# Mechanism responsible for initiating room temperature ferromagnetism and spin polarized current in diluted magnetic oxides

Hsiung Chou[1,*], Yao-Chung Tsao[1], G. D. Dwivedi[1], Cheng-Pang Lin[1], Shih-Jye Sun[2], Hua-Shu Hsu[3], Chun-Chao Huang[4], Chao-Yao Yang[4], Yuan-Chieh Tseng[4], and Fang-Cheng Chou[5]

[1]*Department of Physics, National Sun Yat-Sen University, Kaohsiung 804, Taiwan*

[2]*Department of Applied Physics, National Kaohsiung University, Kaohsiung, Taiwan*

[3]*Department of Applied Physics, National Pingtung University, Pingtung 900, Taiwan*

[4]*Department of Materials Science & Engineering, National Chiao Tung University, Hsinchu, Taiwan*

[5]*Center for Condensed Matter Sciences, National Taiwan University, Taipei 106, Taiwan*

## Abstract

The main obstacles in realizing diluted magnetic oxide (DMO) in spintronics are the unknown electronic structures associated with its high $T_C$ ferromagnetism and spin polarized current and how to manipulate desired electronic structures by fabrication techniques. We demonstrate that fine-tuned electronic structures and band structures can be modified to initiate DMO properties. Interestingly, in the semiconducting state, the doped Co ions and oxygen vacancies contribute non-negligible magnetic moments; and the magnetic coupling between these moments is mediated by the localized carriers via highly spin polarized hopping transport. These results unravel the myth of the origin of spintronic characteristics with desirable electronic states; thereby reopening the door for future applications.

*Correspondent Author: hchou@mail.nsysu.edu.tw

The discovery of diluted magnetic semiconductors (DMSs) [1] and oxides (DMOs) [2] presents an exciting doorway towards realization of spin manipulation [3] in these materials. Most notably, DMOs exhibit ferromagnetic coupling [1,2] and possible spin polarized current (SPC) [3-5] far above room temperature that make them to be possibly used as room temperature operated spintronic devices. However, the mechanisms of magnetic coupling and SPC in DMOs are not clearly understood [6]. As a result, the control of the material and its applications, in terms of reproducibility and spin current control, are difficult. It is well accepted that the bounded magnetic polaron, BMP [2], mechanism initiates the ferromagnetic coupling in insulated materials such as transition metal (TM)-doped ZnO with native oxygen vacancy ($V_O$) concentrations, while Ruderman-Kittel-Kasuya-Yosida (RKKY) [7-9] interaction is responsible for the magnetic coupling in metallic materials such as (Ga, Mn)As or II-VI system [10,11] and degenerated Co and Al co-doped ZnO [12]. For most semiconducting DMOs with a carrier concentration in the range of $10^{16}$~$10^{19}$ cm$^{-3}$, carrier [13-14] or defect [15-17] or grain boundary [18,19] mediated magnetic coupling has been offered as a general explanation, while not providing full details of how these carriers mediate magnetic coupling. Furthermore, doping transition ions in parent DMOs alone does not guarantee magnetic coupling, as certain structural defects are definitely needed [20]. Doped transition ions were found to exhibit very small magnetic moments than the isolated one. Farely *et al.* [21] and Barla *el al.* [22] used XMCD to probe the role of transition ions when partially substituted at Zn sites in ZnO and concluded that the doped transition ions were in paramagnetic state and did not participate in the room temperature ferromagnetic coupling. All room temperature DMO studies show that paramagnetic (PM) and ferromagnetic (FM) components [21-23] coexist simultaneously, indicating that only a portion of Co may participate in magnetic coupling [23]. In some extreme cases, the presence of defects alone is sufficient to generate magnetic coupling even though the cases are rare and the magnetic signals are quite weak and without hysteresis [18,24]. This indicates that the emergence of ferromagnetic coupling in a TM-doped ZnO DMO is intimately associated with the details of the electronic structure, such as the relative positions of the transition-ion's subbands and $V_O$ states to the conduction band. However, up to now, the mechanism of DMS revealed by both the theoretical calculations [25-29] and experimental measurements have been controversial. The locations of TM-subbands have been predicted to reside within the conduction band or to be within the bandgap for either very far from or near to Fermi levels. The defect states have been found to be either deep or shallow donor or acceptor states. Moreover, researchers paying more attention on the emergence of high $T_C$ ferromagnetism, the spin coherent characteristics of transport carriers, the major issue for diluted magnetic materials research and future applications, generally remained unexplored [23]. Due to the non-reproducible fabrication procedure in accompany with unknown mechanism, the highly desired spin manipulating potential in DMO materials is still uncontrollable. Accordingly, one of the main challenges in modern oxide spintronics is to regain total control of the material and manipulation of spins. This requires developing a detailed understanding of the intrinsic mechanism and a reproducible production technique.

In this study, we concentrate on the transition ion doped ZnO system because its DMO characteristics are much easier to be observed. Fine tuning of the electronic structure has been achieved by controlling both grain sizes and $V_O$ on Co-doped ZnO diluted magnetic oxide such that the correlation of electronic structure to spintronic characteristics can be clarified. In the fine-tuning procedure, the empty Co-3d-$t_{2g}\downarrow$ subband can be shifted to either above or below the conduction band minimum (CBM) and the DMO characteristics, those of room temperature magnetic coupling and SPC, can be initiated by various methods. For the former case, corresponding to the large grains and low $V_O$ concentration samples, injection of adequate carriers may raise the Fermi level to the empty Co-3d-$t_{2g}\downarrow$ subband and trigger the RKKY coupling. For the latter case, corresponding to the small grain and high $V_O$ concentration samples, introducing effective defects that overlap with the empty Co-3d-$t_{2g}\downarrow$ subband around the Fermi level may provide hopping carriers which initiate a strong *p-d* coupling. This mechanism, based on n-type ZnO DMO, can also be adopted and applied to explain p-type DMOs [23], but some of the $d^0$ or GB DMOs [17-19] which strongly related with the complex defect structures are not included in this study. These results unravel the mystery of the origin of spintronic characteristics in DMOs with desirable electronic states, thereby restoring control of the specific electronic structure and the manipulation of SPC.

A batch of $Zn_{0.95}Co_{0.05}O_{1-\delta}$ (CZO) films were grown on glass and fused quartz substrates by standard magnetron sputtering in reducing atmosphere, a mixture of Ar and 1~30% of $H_2$ (denoted as #% such that CZO films grown at a specific $H_2$% are written as CZO-#% throughout this paper). Films were grown at room temperature, unlike other high temperature processing conditions, to reduce the possibility of hydrogen doping on films. We found that Co fully incorporates into the Zn sites and no hydrogen ions contained in films.

Since the ferromagnetic coupling in ZnO DMO is the interplay between the electronic structures of the ZnO matrix, the magnetic ions' subbands, and the $V_O$ states, a MCD spectroscope is used to reveal the fine electronic structure. Figure 1 shows the optical density (OD) spectra (pink lines), the 1st derivative of OD (red dots) and the MCD spectra (blue dots) of CZO-1% and -10% samples, respectively. The peaks observed in first derivative of OD spectra defines the location of absorption edge of respective system. The CZO-1% sample has an obvious absorption edge of 3.45 eV, Figure 1(a-1), similar to the conduction-valence band gap of pure ZnO. Its MCD spectrum exhibits several peaks around 2, 3.1, 3.4 and 3.67 eV. The 2 eV peak is the spin-orbit split $^4A_2 \rightarrow {}^4T_1(P)$, or *d-d* transition, due to the crystal field of $Co^{2+}$ subbands being doped substitutionally in the ZnO lattice [30,31]. The 3.4 eV peaks have many possible sources [23]. This kind of splitting was not observed in pure ZnO [32]. If we take the 3.67 eV peak as the MCD main transition, then the change in its absorption intensity is clearly much smaller than that of the 3.45 eV edge and therefore is very difficult to be identified in the OD spectrum. In contrast, the 3.45 eV optical absorption edge is not observed in the MCD spectrum and must not be a transition due to spin imbalance. It is reasonable to assign the 3.45 eV absorption edge as the optical excitation between the valence and conduction band of ZnO (the optical band gap $E_{g\text{-MCD}}$) and the other 3.67 eV MCD peak as a spin polarized MCD main transition, $E_{MCD}$. Similar analysis are carried out on CZO-10% sample as shown in Figure 1(b), and also applied to all samples. In Figure 1(b), it is surprising that the $E_{MCD}$ of CZO-10% sample is lowered to 3.47 eV, while the $E_{g\text{-MCD}}$ is raised to 3.56 eV. The optical band-gaps of all samples, estimated from the transmittance spectra, optical absorption spectra and the $E_{MCD}$, are plotted in green solid inverse-triangles, red solid triangles and solid blue squares, respectively in Figure 2(a). The obvious crossover feature of $E_{g\text{-MCD}}$ and $E_{MCD}$ indicates a very important essential change in the local band structure that is found strongly correlated with the mechanism of DMO and suggests a method to control the band structure by manipulating the material preparation process.

Identifying the $E_{MCD}$ is important in disclosing the band structure detail. Although band calculations show controversial results [25-29,33], one common feature is that the Co-3d subbands remain very localized (similar to its atomic orbitals) and can be divided into four subsets: $e_g\uparrow$, $t_{2g}\uparrow$, $e_g\downarrow$ and $t_{2g}\downarrow$ for a tetrahedral coordination to hybridize with O-2$p$ and Zn-4$s$ orbitals. At ground state, Co-3d-$e_g\uparrow$, -$t_{2g}\uparrow$ and -$e_g\downarrow$ subbands are filled, while -$t_{2g}\downarrow$ subband is empty. Among many diverse calculation results, Tsai's calculation [33] supports our observation qualitatively. Tsai *et al.* found that, without the presence of oxygen vacancies, the empty Co-3d $t_{2g}\downarrow$ band is located inside the conduction band while the $e_g\downarrow$ is located right below the CBM. Because the conductivity of Co-doped ZnO without $V_O$ is very close to zero (Figure 2(b)), the Fermi energy, $E_F$, is identified as being located between the Co-3d $e_g\downarrow$ band and CBM. Because the partial density of states of the empty Co-3d $t_{2g}\downarrow$ band is at least 20 times larger than the Zn-4s conduction band, the 3.67 eV $E_{MCD}$ must be the transition between the valence band maximum (VBM) and the empty Co-3d $t_{2g}\downarrow$ band. The schematic band structure of the CZO-1% sample is plotted in Figure 3(a), where the empty Co-3d $t_{2g}\downarrow$ band is 220 meV higher than the CBM. Similarly, the 3.1eV excitation MCD peak in the spin-up side, which is not observable in a pure ZnO sample, can be regarded as the charge excitation process from a localized band, Co-3d-$t_{2g}\uparrow$, located inside the band gap to the CBM.

The schematic band structure can be examined by the electric field-biased-XMCD on the samples consisting large grains with low $V_O$. We found that the injection of carrier (equivalent to raising the Fermi level) makes an obvious change in Co-L2,3 XMCD peaks [34]. This phenomenon indicates that the empty Co-3d $t_{2g}\downarrow$ band is located inside the conduction band and the XMCD L2,3 peaks are changed when the Fermi level is raised up to the Co-3d $t_{2g}\downarrow$ band. Conversely, the XMCD L2,3 peaks do not change when the Fermi level is lowered by the extraction of carriers. The present results can be compared with Li's experiment [20], where a single crystal CZO film was grown on a ZnO single crystal substrate and the film could be regarded as free from $V_O$. Li *et al.* found that all doped Co ions, as long as the Co concentration was less than 16%, acted paramagnetically. This finding is consistent with our own, as the Fermi level of their samples was located between the CBM and the filled Co-3d-$e_g\downarrow$ subbands.

Interestingly, the oxygen XMCD of our samples consisting of $V_O$ is non-zero in any conditions. This strongly suggests that part of the $V_O$ provide magnetic moments. In CZO-1%, therefore, both $V_O$ and Co ions provide magnetic moments.

By decreasing grain sizes and increasing $V_O$, the CZO-10% exhibit dramatic changes in the $E_{g-MCD}$ and $E_{MCD}$, (Figure 1(b) and 2). The $E_{g-MCD}$ increased to 3.56 eV, while the $E_{MCD}$ decreased to 3.47 eV. The empty Co-3d-$t_{2g}\downarrow$ subband now resides 90 meV below the CBM (Figure 3(b)). Because the concentration of the Co dopant is fixed, the more disordered the structure, the more indistinct the empty Co-3d-$t_{2g}\downarrow$ subband and the localized states; as a result, their boundaries become indistinguishable and they hybridize along the indistinct boundaries and overlap with the Fermi level (Figure 3(b)). This picture is confirmed by Tsai's calculation that $V_O$ reduces the Co-3d subbands' energy and the Co-3d subband will polarize and hybridize O-2p states. The electric field-biased-XMCD measurements find that regardless of whether the carrier is injected into or extracted from the sample, both XMCD signals of Co and O show profound fluctuations. The $V_O$ localized states and the empty Co-3d-$t_{2g}\downarrow$ subband overlap around the Fermi level. Therefore, a *p-d* coupling is the mechanism of the observed ferromagnetic coupling.

The trend of magnetism estimated by MCD has also been confirmed by SQUID measurements, denoted as black center dotted triangle in Figure 2(b). The CZO-10% film exhibits the largest saturated magnetization of 13 μemu/cm$^3$, which is equal to 0.044 $\mu_B$/Co if the magnetic moments are attributed to the all doped Co ions, while the CZO-30% is only 5 μemu/cm$^3$. The original magnetization curves are dominated by a strong paramagnetic background, with 0.53 and 0.044 $\mu_B$/Co for paramagnetic and ferromagnetic magnetization. The average magnetic moment per Co ions came up to a value of 0.57 $\mu_B$ which is much smaller than the moment 3 $\mu_B$ of isolated Co ions. This phenomena has been observed by many groups [20-23]. For example, the single crystal $Zn_{1-x}Co_xO$ films grown by Li *et al.* [20] had a uniform distribution of Co, confirmed by a 3D atom probe tomography, and showed an average moment of 0.09 $\mu_B$/Co for x=0.1. Since the structural defect of their single crystal film can be neglected, this low average magnetic moment of Co implies a hybridization effect between Co-3*d* and O-2*p* and Zn-4*s* orbitals in such a way, which leads to a low average moment of Co. This is supported by Tsai's calculations [33] and he pointed out that the Co-3*d* band is strongly hybridized with O-2*p* and Zn-4*s* orbitals.

The band diagrams in Figure 3 reflect the conceptual part of the complete electronic structure in momentum space. A picture in real space is built to depict how the magnetic coupling and the hopping conduction take place. Transport measurements reveal that both the localized radius, $r_{VRH}$, and the hopping radius, $r_R$, of carriers around the shallow $V_O$ states increase with $V_O$ concentration. For the CZO-10% sample, for example, the $r_{VRH}$ and $r_R$ are 0.8 nm and 1 nm, respectively. Four types of spheres according to the $V_O$ conditions and magnetic polaron radii can be visualized (see the lower panel of Figure 4), where disks rather than spheres were used for graphic clarity. The moment of the $V_O$ may strongly couple with the localized spins within a small $r_{VRH}$ via localized carriers to form a magnetic polaron sphere; or the coupling can only happen within an inner sphere if the $r_{VRH}$ is large, as proposed by the concentrate bounded magnetic polaron [35]. These couplings are strong enough to overcome the spatial anisotropy and thermal perturbation and to align all these moments in the effective sphere into the same direction. The magnetic coupling between adjacent spheres is accomplished by hopping carriers if the adjacent inner spheres are close to each other within the range of $r_R$. For the CZO-1% sample, shown in Figure 4(a), very few spheres are formed and are separated by a distance much longer than $r_R$ such that interactions between adjacent spheres are impossible. By increasing the concentration of $V_O$, more magnetic polaron spheres are formed and some of them may near to one another and have chance to coupled together to form ferromagnetic clusters. The magnetic strength and the saturated magnetic signal at low $V_O$ concentrations are increased proportional to the number of spheres, which are themselves proportional to the $V_O$ concentration. When the concentration of $V_O$ is further increased, a percolation of magnetic polaron spheres throughout the entire material contributes ferromagnetic signal, as in the case of the CZO-10% sample in Figure 4(b). For the CZO-30% sample, massive $V_O$ were created and the $E_{g\text{-MCD}}$, $E_{MCD}$ and the excitation energy of shallow donor state shift to 3.6(~3.7) eV, 3.5 eV and 80 meV, respectively, indicating the empty Co-3d-$t_{2g}\downarrow$ band has a nearly perfect match with the $V_O$ states. A massive charge transfers from the $V_O$ to the empty Co-3d-$t_{2g}\downarrow$ subband, and the magnetic moment of spheres will decrease due to the flood effect, where the minority moments cancel out the majority ones. At this stage, the number of coupled spheres increases while the moments of each sphere decreases. The competition, between these two effects, induces the maximum saturated magnetization for the CZO-10% sample. With further increases in $V_O$, the flood effect brings the saturated magnetization down, such as in the CZO-30% sample. The present model have been applied to other systems [12,23,36] to depict the underlying mechanism. For details, please see the supplementary material.

This study clarifies the distinct DMS mechanism in a ZnO system in a semiconductor state where the magnetic moments can be mediated by the hopping carrier when the localized hopping states and the magnetic subbands hybridize around the Fermi level. The number of localized hopping states is very limited for native materials and can be increased artificially. Previously, it was difficult to situate the magnetic subbands at a proper energy level. This work achieved this by manipulating particle size and the concentration of artificial localized states to induce a disordered system. Once both the localized hopping states and the magnetic subbands shift near Fermi level, the strong hybridization between them will trigger magnetic coupling. Moreover, when the percolation limit is reached, a fully spin polarized hopping transport will govern the conduction property and the material will be, then, a highly conductive half-semiconductor.

## Acknowledgements

The authors would like to acknowledge the great help from Professor Wei-Feng Tsai and Professor Michael Chiang of National Sun Yat-sen University, Professor Cheng-Hsuan Chen of the Center for Condensed Matter Sciences, National Taiwan University, Professor Mark Blamire, Dr. Zoe Barber and Jason Robinson of University of Cambridge, UK. This project is supported by the Ministry of Science and Technology of Taiwan under Grant No. NSC-102-2112-M-110-MY3.

**Figure 1:** The MCD measurement curves and analysis for (a) CZO-1%, and for (b) CZO-10% sample. The optical density, the first derivative of optical density and the MCD spectra are plotted.

**Figure 2:** Summary of important data as a function of $H_2$%. (a) The Optical bandgaps and the MCD main transition. (b)The room temperature conductivities, MCD strengths, SQUID measurements and the contributions of VRH and TE mechanism at room temperature for the CZO-30% sample.

**Figure 3:** DMO mechanism in momentum space plotted as conceptual band structures of the samples (a) CZO-1%, and (b) CZO-10%.

**Figure 4:** DMO mechanism in real space. Four types of basic coupling spheres are shown in the lower panel as disks rather than spheres for graphic clarity, and correspond to how the sphere interacts with magnetic moments. When $V_O$ are few, only isolated sphere scattered within samples, (a). While $V_O$ increases, a few spheres may couple to form superparamagnetic domains, shown as the clusters on the right half of (b). A percolation path, as shown on the left side of (b), is formed when the films contain a very high $V_O$ concentration, and ferromagnetic strips appear.

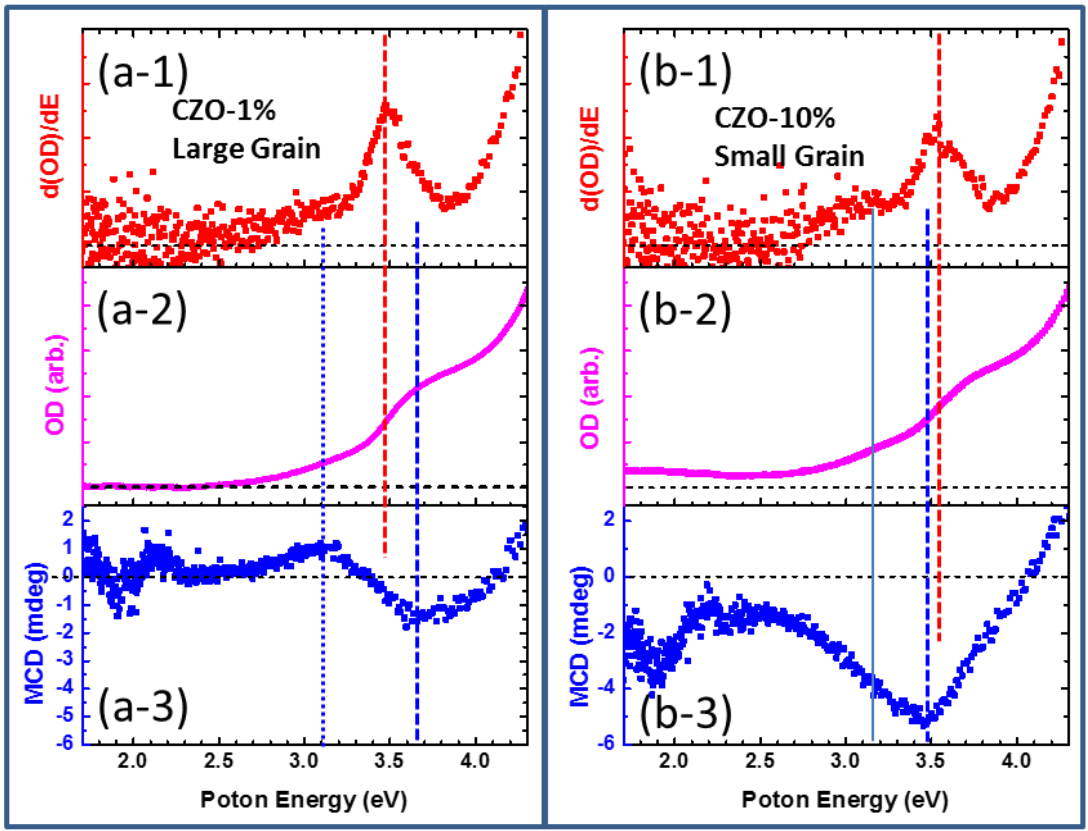

Figure 1

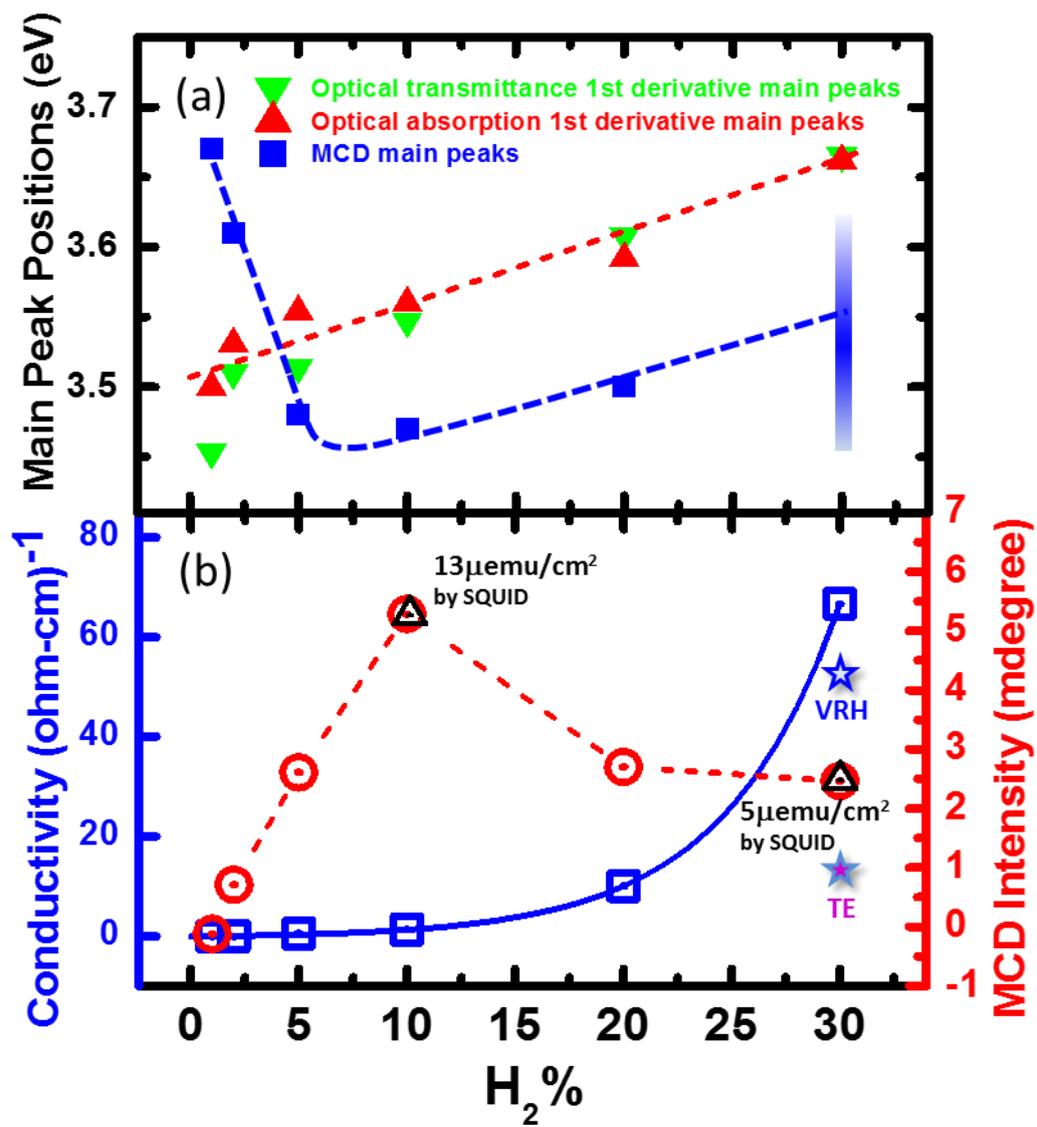

Figure 2

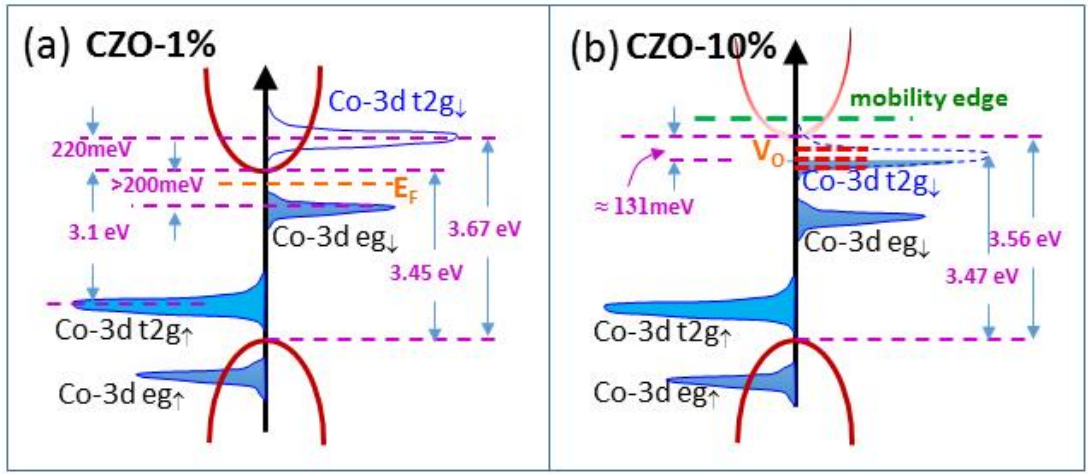

**Figure 3**

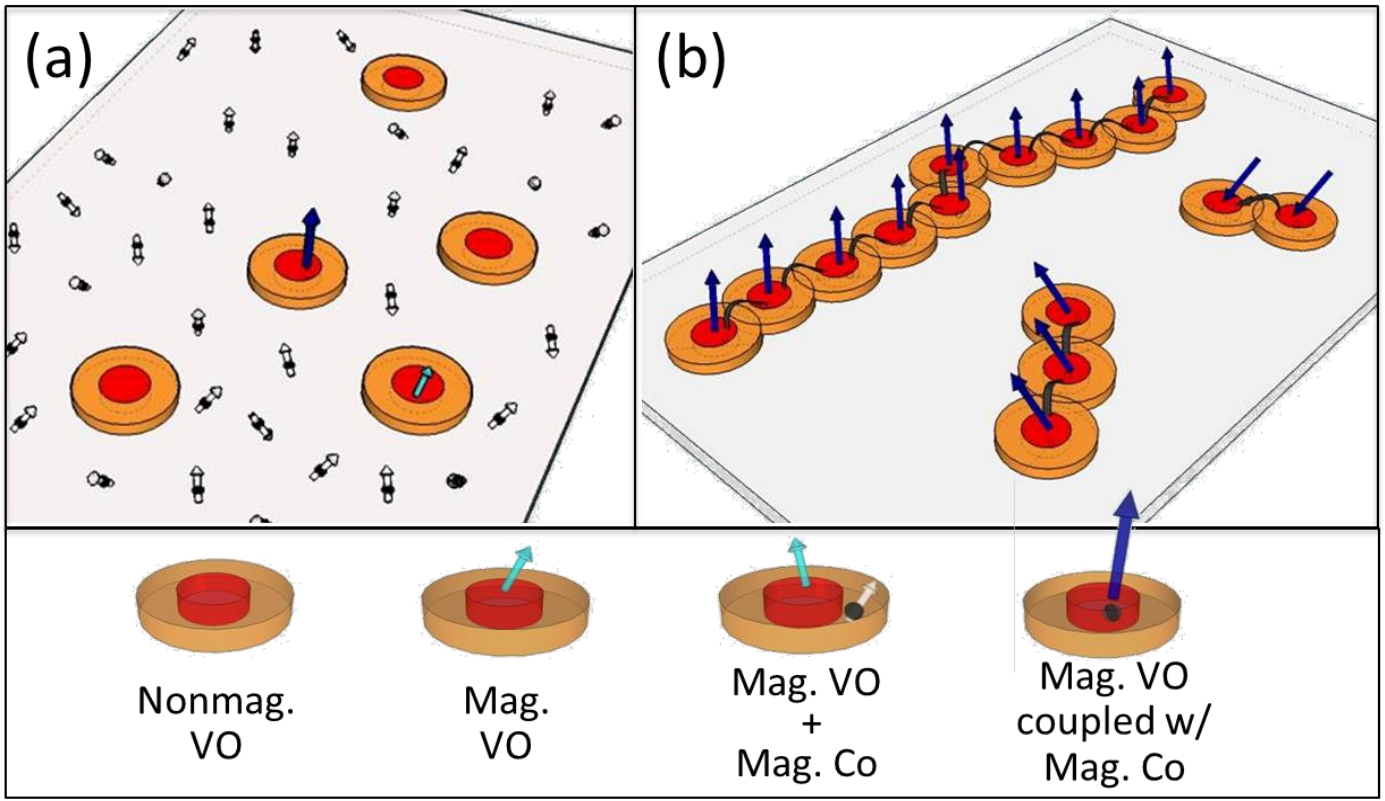

**Figure 4**